\documentclass[11pt,twoside]{article}

\usepackage{asp2006}
\usepackage{epsf}
\usepackage{psfig}
\usepackage{lscape}

\begin{document}
\title{High Accuracy Imaging Polarimetry with NICMOS\altaffilmark{1}}   
\author{D. Batcheldor,\altaffilmark{2} G. Schneider,\altaffilmark{3} D. C. Hines,\altaffilmark{4} G. D. Schmidt,\altaffilmark{3} 
D. J. Axon,\altaffilmark{5} A. Robinson,\altaffilmark{5} W. Sparks\altaffilmark{6} \& C. Tadhunter\altaffilmark{7}}
\altaffiltext{1}{Based on observations made with the NASA/ESA Hubble Space Telescope obtained at the 
Space Telescope Science Institute, which is operated by the Association of Universities for Research in Astronomy, Incorporated, under 
NASA contract NAS 5-26555.}
\altaffiltext{2}{Center for Imaging Science, Rochester Institute of Technology, 54 Lomb Memorial Drive, Rochester, NY, 14623, USA}
\altaffiltext{3}{Steward Observatory, The University of Arizona, 933 N. Cherry Avenue, Tucson, AZ, 85721, USA}
\altaffiltext{4}{Space Science Institute, 4750 Walnut Street, Suite 205, Boulder, CO 80301, USA}
\altaffiltext{5}{Department of Physics, Rochester Institute of Technology, 54 Lomb Memorial Drive, Rochester, NY, 14623, USA}
\altaffiltext{6}{Space Telescope Science Institute, 3700 San Martin Drive, Baltimore, MD, 21218, USA}
\altaffiltext{7}{Department of Physics \& Astronomy, University of Sheffield, Sheffield, S3 7RH, UK}

\begin{abstract} 
The ability of NICMOS to perform high accuracy polarimetry is currently hampered by an uncalibrated residual instrumental polarization 
at a level of $1.2-1.5\%$. To better quantify and characterize this residual we obtained observations of three polarimetric standard 
stars at three separate space-craft roll angles. Combined with archival data, these observations were used to characterize the residual 
instrumental polarization to enable NICMOS to reach its full polarimetric potential. Using these data, we calculate values of the parallel 
transmission coefficients that reproduce the ground-based results for the polarimetric standards. The uncertainties associated with the 
parallel transmission coefficients, a result of the photometric repeatability of the observations, dominate the accuracy of $p$ and 
$\theta$. However, the new coefficients now enable imaging polarimetry of targets with $p\approx1.0\%$ at an accuracy of 
$\pm0.6\%$ and $\pm15\deg$.
\end{abstract}


\section{Introduction}

The Near Infrared Camera and Multi-Object Spectrometer (NICMOS) is the {\it only} near infrared (NIR) instrument capable of the high resolution, 
high fidelity polarimetry needed to examine the scattering geometry and materials in many types of astronomical objects. However, there have been 
recent reports of a systematic residual instrumental polarization of $\sim1.5\%$ \citep{2005AJ....129.1625U} that would seriously compromise 
studies of objects such as active galactic nuclei and regions within circumstellar protoplanetary disks; they typically exhibit polarizations 
of $<5\%$. A comprehensive study of this residual polarization was carried out by \citet[B06]{2006PASP..118..642B}, who found a residual excess 
polarization of $\approx1.2\%$ in the NIC2 camera. 

Prior to this work, observations of only one polarized and one unpolarized standard have been made, each at only two celestial 
orientation angles. Therefore, until now, it has been impossible to entirely characterize the NICMOS residual polarization. Polarization 
measurements at three well separated roll angles are necessary to remove the dependence of the measured Stokes parameters on the relative 
transmission of each of the filters. We report here the findings of a calibration program that collected such data. 

\begin{table}[t]
\caption{Details of Polarimetric Standards \label{tab:stand}}
\smallskip
\begin{center}
{\small
\begin{tabular}{lccccccc}
\tableline
\noalign{\smallskip}
Target & Type & $m_v$ & $m_j$ & $m_k$ & 0.55\micron & 2.00\micron & 2.04\micron \\
       &      &       &       &       &\multicolumn{3}{c}{$(p\%,\theta)$}       \\
\noalign{\smallskip}
\tableline
\noalign{\smallskip}
VR 84c (WKK F7) & B5V   & 10.3 & 9.34 & 9.13 & 5.98,118\deg & 1.25,126\deg & 1.19,126\deg \\   
VSS VIII-13     & K1III & 12.7 & 9.65 & 8.68 & 2.25,102\deg & 0.91,102\deg & 0.86,102\deg \\
HD 331891       & A4III & 9.3  & 8.80 & 8.72 & 0.04,n/a     & ...          & ...    \\
\noalign{\smallskip}
\tableline
\end{tabular}
}
\end{center}
\end{table}

\section{Observations and Data Reduction}

The observations addressed the NICMOS residual polarization and several other key issues, including possible dependence with detector 
quadrants, source color dependence, the inter-pixel response functions (IPRFs) and latent image persistence. All orbits were scheduled 
so not to be impacted by the effects of the South Atlantic Anomaly. Persistence can be mitigated with an effective image dithering strategy. 
However, a precise dithering pattern has to be chosen to avoid image raster overlaps of the PSF diffraction spikes. The observations 
were dithered in each detector quadrant to circumvent the effects of the IPRFs and sample around defective pixels. The pointing offsets 
covered a phase spacing of -1/3, 0 and +1/3 pixels to best tile the IPRFs for the dither pattern employed. The dither steps were 
sufficiently large so that the $2^{\rm nd}$ Airy PSF maxima from each dither (0\farcs58, 7.6 pixels), did not overlap from position to position. 

The three polarimetric standard stars presented in Table~\ref{tab:stand} were observed at three separate space craft 
roll angles. This provides the data for a unique determination of any residual instrumental polarization relative to the equatorial reference 
frame. The technique allows the polarimetric analysis, derived from the three polarizers at each of the field orientations, to be decoupled and 
checked for consistency. For comparison with the NIC2 data, the ground-based polarization measurements are corrected to 2.00\micron~ using the 
known wavelength dependence of interstellar polarization \citep{1975ApJ...196..261S}. 

It is essential to inspect the fidelity of each pointed observation. Instrumental sources of variation in the photometry, not directly resulting 
from the intrinsic polarization of the target, must be mitigated. Therefore, radially increasing aperture photometry was 
performed at each pointing and the encircled count rates extracted. Radial profiles that showed more than a $2\sigma$ deviation from 
the average profile were removed from further study. 

\begin{figure}[t]
\plotone{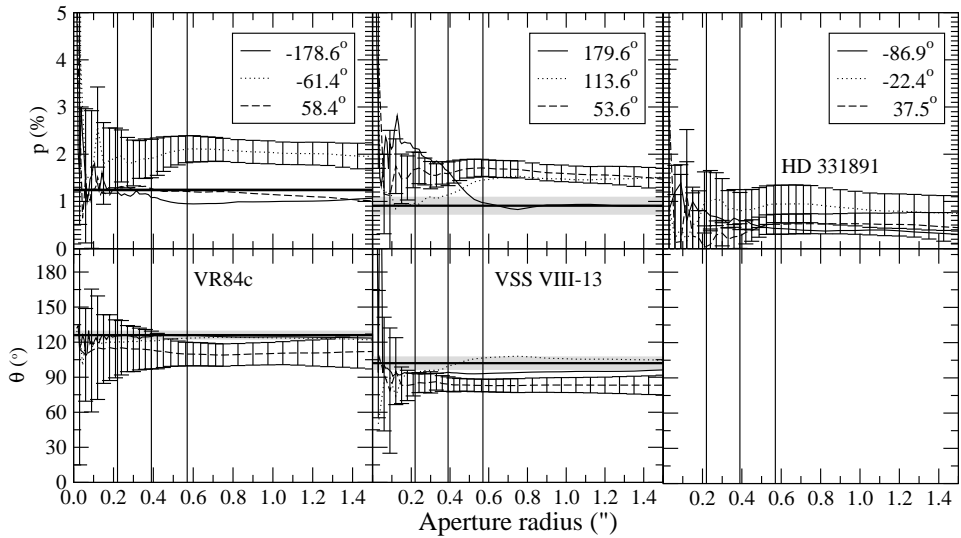}
\caption[Previous Calibration]{
The results using the previous calibration. The thick solid lines show the ground based results and the grey shaded areas show the 
uncertainties associated with those ground-based results.  Error bars are only shown for the cases where they do not overlap the 
ground based results. The vertical solid lines mark the radii of the $1^{\rm st}, 2^{\rm nd}$ and $3^{\rm rd}$ dark minima of the PSF. 
Each line marks the three celestial field orientations for each target indicated.
}\label{fig:oldcal}
\end{figure}

The determination of $p$ and the orientation of polarization ($\theta$) is addressed for NICMOS by \citet[HSS]{2000PASP..112..983H} and 
B06. In order to assess the previous polarimetric accuracy, we have applied the coefficients derived by HSS to 
the new NIC2 calibration data. The results are presented in Figure~\ref{fig:oldcal}, which demonstrates that at only a few roll angles 
does the previous calibration reproduce the ground-based results. 

\section{The Calibration}

Since we have observed each standard at multiple space craft roll angles, the uncertainties due to the relative filter transmissions 
can be removed; each of the three polarizers, separately, was used to collect all the necessary polarimetric data. With these unknowns 
nullified by the multiple orientations, the only remaining explanation for these residuals (beyond intrinsically smaller uncertainties 
in the ground-based calibration) is an instrumental polarization ($p_i$). As $p_i$ is fixed to the space-craft $(Q,U)_i$ are in the 
instrumental frame; $p_i$ must be determined in Stokes space. Therefore, we determine the actual values and uncertainties of $Q$ and 
$U$ $(Q,U)_{ac}$, given the ground-based results, and determine $(Q,U)_i$ assuming $(Q,U)_{i} = (Q,U)_{ob} - (Q,U)_{ac}$. Using this 
method we find the average instrumental polarization is $0.6\pm0.1\%$ at $116\pm15$\deg. 

\begin{figure}[t]
\plotone{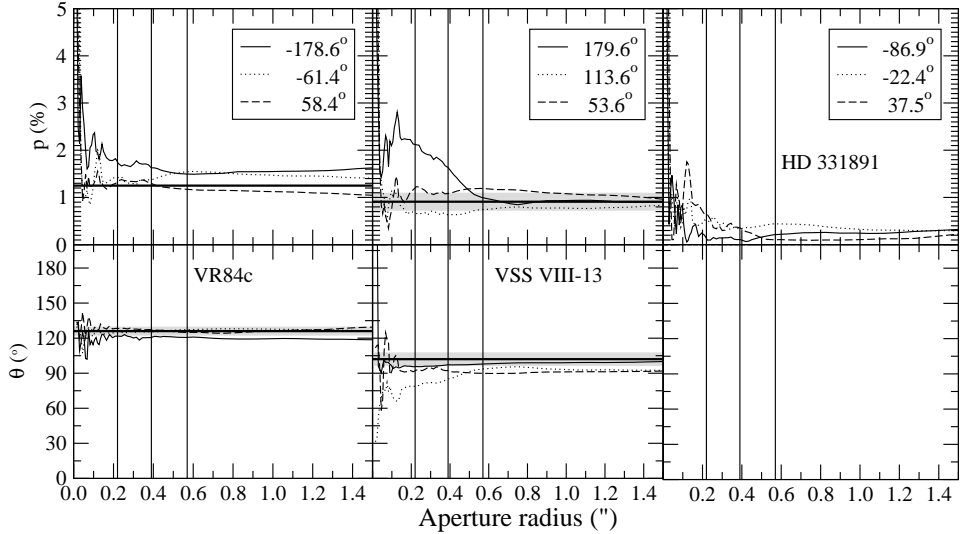}
\caption[The New Calibration]{
The results from the new calibration. The same convention is used in Figure~\ref{fig:oldcal}. All results are now consistent with 
the ground-based data.
}\label{fig:newcal}
\end{figure}

\begin{table}[ht]
\caption{The New Polarimetric coefficients\label{tab:coeffs}}
\smallskip
\begin{center}
{\small
\begin{tabular}{lcccc}
\tableline
\noalign{\smallskip}
Polarizer & $\phi_k$ (\deg) & $\eta_k$ & $l_k$ & $t_k$ \\
\noalign{\smallskip}
\tableline
\noalign{\smallskip}
POL0L   & 8.84   & 0.7313 & 0.1552 & $0.882\pm0.008$ \\
POL120L & 131.42 & 0.6288 & 0.2279 & $0.837\pm0.004$ \\
POL240L & 248.18 & 0.8738 & 0.0673 & 0.9667 \\
\noalign{\smallskip}
\tableline
\end{tabular}
}
\end{center}
\end{table}

With $p_i$ determined (and subtracted), the parallel transmission coefficients ($t_k$) that reproduce 
the ground-based results can then be determined numerically. The parallel transmission for POL240L ($t_{240}$) is held 
at 0.9667 as per HSS. B06 also performed this process on archival data, so we include the values of $t_k$ derived by 
those authors. Table~\ref{tab:coeffs} presents the newly derived polarimetric coefficients from this study. There are 13 
independent determinations of the coefficients from each orbit (nine new, four archival), the standard deviation 
of which is used to define the uncertainty in the coefficients about the mean. These uncertainties are propagated 
through the calculation of $p$ and $\theta$. As a check, the new coefficients have been applied to the calibration 
data and presented in Figure~\ref{fig:newcal}. This calibration gives results consistent with the ground-based data.

\section{Discussions}\label{dis}

There are many potential (known and conjectured) sources for the estimated errors on $p$ and $\theta$. However, the uncertainties 
in the derived transmission coefficients dominate the statistical uncertainties from the photometry. Could these uncertainties be 
responsible for the observed instrumental polarization? A more accurate way to determine $p_i$ requires observations of an unpolarized 
flux standard across the whole detector focal plane; this would map the two-dimensional relative transmission differences of the single 
polarizers. If there is a relative transmission across the focal plane, then this could mimic an instrumental polarization. Therefore, 
it is appropriate to present the 0.6\% value as an instrumental upper limit; $p_i$ could be zero. 

When compared with the previous calibration, it can be seen that $t_{120}$ is consistent; $t_0$ has been changed by 0.5\%. 
Therefore, applying the new values back to previous studies of targets with $p>5\%$ will not significantly alter 
the results. However, when applied to targets with $p<5\%$, the new value of $t_{120}$ results in a very significant improvement 
in polarimetric accuracy.

We recommend specific observing and data analysis strategies for spatially unresolved, and resolved, targets. For a spatially resolved 
target (without a bright point source) observers should bin their data to $3\times3$ pixels (the diffraction limit of the instrument). 
Dithering can be used to increase spatial resolution and sample around bad pixels. However, high signal to noise ratio 
data is still critical in determining $p$ and $\theta$ to high accuracy. The case of a spatially unresolved source is more complicated. 
Due to uncorrectable bad pixels, the dithered data must be inspected to identify deviant profiles. This must be done by executing a 
large number of pointed dithers (with a step size greater than the PSF to avoid persistence). Encircled energy profiles should 
be derived independently from each observation. A sigma clip should then be used to determine 
whether an individual profile is deviant or not. The greater number of dithers used, the more accurate the flagging of deviant profiles 
will be. For small numbers of dither points (not optimally recommended) medianing can be used to coarsely reject outliers. Outside an 
aperture of radius 0\farcs57, the PSF affects disappear. Inside this radius, the uncertainties in $p$ and $\theta$ rise rapidly. 
Observers must weigh the required polarimetric accuracy to the required resolution. 

\section{Conclusions}

Using the multi-roll technique, we have placed an upper limit to the NIC2 instrumental polarization of 0.6\%. The instrumental 
polarization has been subtracted from the target data in the $Q,U$ plane. New parallel transmission coefficients were then derived 
by forcing $p$ and $\theta$ to the same values of the calibration standards. The new coefficients are such a minor change from the 
previous that it does not warrant the re-analysis of previous NIC2 imaging polarimetry data of highly polarized ($p>5\%$) targets. 
Importantly, this improved calibration opens a new domain for observational investigations with {\it HST} by enabling very high 
precision polarimetry of intrinsically low ($p<5\%$) polarization sources.

\acknowledgments

Support for Proposal number HST-GO-10839.01-A was provided by NASA through a grant from the Space Telescope Science Institute, 
which is operated by the Association of Universities for Research in Astronomy, Incorporated, under NASA contract NAS5-26555.

\end{document}